\documentstyle{proceeding}
\input epsf.tex

 \mathcode`*="002A
\newbox\grsign \setbox\grsign=\hbox{$>$} \newdimen\grdimen \grdimen=\ht\grsign
\newbox\simlessbox \newbox\simgreatbox \newbox\simpropbox
\setbox\simgreatbox=\hbox{\raise.5ex\hbox{$>$}\llap
     {\lower.5ex\hbox{$\sim$}}}\ht1=\grdimen\dp1=0pt
\setbox\simlessbox=\hbox{\raise.5ex\hbox{$<$}\llap
     {\lower.5ex\hbox{$\sim$}}}\ht2=\grdimen\dp2=0pt
\setbox\simpropbox=\hbox{\raise.5ex\hbox{$\propto$}\llap
     {\lower.5ex\hbox{$\sim$}}}\ht2=\grdimen\dp2=0pt
\def\simgreat{\mathrel{\copy\simgreatbox}}
\def\simless{\mathrel{\copy\simlessbox}}

\begin{document}
\title{The nature of the X-ray source in NGC 4151}
\author{P. Magdziarz\inst{1} \and A.A. Zdziarski\inst{2}}
\institute{Jagiellonian University, Astronomical Observatory, Orla 171, 
 30-244 Cracow, Poland  
\and  N. Copernicus Astronomical Center, Bartycka 18, 00-716 Warsaw, Poland}  
\maketitle
\begin{abstract}

Analysis of broad-band X/$\gamma$-ray spectra of NGC 4151 shows 
that the data are well modelled with an intrinsic spectrum due to 
thermal Comptonization with temperature $\sim$ 50 keV and X-ray 
spectral index $\alpha\sim$ 0.4--0.7. The variable X-ray spectrum pivots
at $\sim$ 100 keV, which is consistent with the observed approximatly 
constant $\gamma$-ray spectrum. The observed UV/X-ray correlation can be 
explained by two specific models with reprocessing of X-rays by cold matter. 
The first one is based on reemission of the X-ray flux absorbed by clouds 
in the line of sight. The second assumes reprocessing of X-rays and 
$\gamma$-rays by a cold accretion disk with a dissipative patchy corona. 

\end{abstract}

\section{Introduction}

NGC 4151 is a nearby Syfert 1.5 galaxy. Its X-ray spectrum is  
highly variable in both the 2--10 keV flux and the 2--10 keV 
spectral index (e.g. Yaqoob et al.\ 1993). The X-ray flux shows a good 
correlation with the UV flux (Perola et al.\ 1986). The spectrum is consistent 
with optically-thick thermal Comptonization rather than highly-relativistic
optically-thin one, which is characteristic for less variable spectra of 
Syfert 1's. We consider here the origin of the variability of NGC 4151. 
 
\section{Spectral variability}

Zdziarski, Johnson \& Magdziarz (1996) analyzed the broad-band X/$\gamma$-ray 
spectra of NGC 4151 from  
contemporaneous observations (Fig.~1), and they found that the data are 
well modelled with an intrinsic spectrum due to thermal Comptonization 
(Titarchuk \& Mastichiadis 1994). The X-ray energy spectral 
index changes from $\alpha\sim$ 0.4 to 0.7, and the temperature stays at 
$\sim$ 50 keV. The spectra show no Compton reflection component.  
Other observations by OSSE up to Dec. 1993 also show that the observed 
spectra and fluxes are roughly constant and can be described by thermal 
Comptonization at $kT \simeq 60$ keV. The pattern of spectral variability 
can be described by varying X-ray power law pivoting at $\sim$ 100 keV 
(cf. Fig~1). 
The spectrum breaks at that energy permitting the $\gamma$-ray flux  
to stay approximately constant. 

We also reanalyzed X-ray observations from {\it IUE} /{\it EXOSAT} campaigns 
in 1983 Nov.\ 7--19 and 1984 Dec.\ 16--1985 Jan.\ 2 (Perola et al.\ 1986). 
Refitting of the {\it EXOSAT} data with a power law and a constant soft 
excess component resulted in a correlation close to linear between  
the absorption-corrected $EF_E$ at 5 keV and at 8.5 eV (from $IUE$ 
de-reddened observations; Ulrich et al.\ 1991), which values are comparable.  

\begin{figure}
\begin{center}
\leavevmode 
\epsfysize=5cm\epsfbox[8 62 544 424]{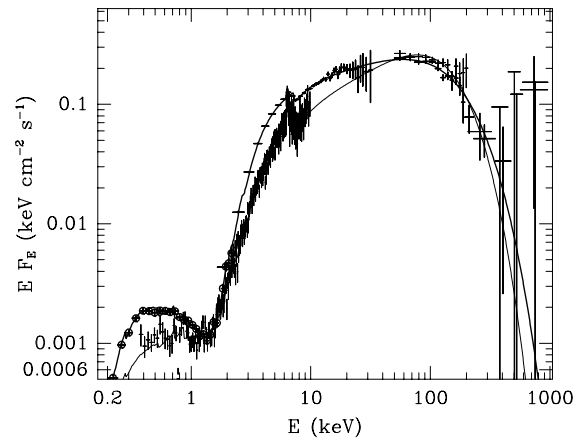}
\end{center}
\caption[]{Thick symbols show the spectrum observed by {\it GRO}/OSSE, 1991 
June 29--July 12 
and by {\it Ginga\/} (Yaqoob et al.\ 1993) and {\it ROSAT\/} (marked by 
circles, Warwick, Done \& Smith 1995), 1991 May 31--June 1. 
Thin symbols show the spectrum observed by OSSE, 1993 May 24--31, and by 
{\it ASCA}, 1993 May 25. The spectra are described by thermal  
Comptonization models with $kT=66$ keV, $\alpha=0.65$ and $kT=47$ keV, 
$\alpha=0.44$ (thick and thin solid curves respectively) absorbed by an 
ionized medium, and with addition of a soft excess (Zdziarski, Johnson
\& Magdziarz 1996).}
\end{figure}

\begin{figure}
\begin{center}
\leavevmode 
\epsfysize=5cm\epsfbox[117 62 437 424]{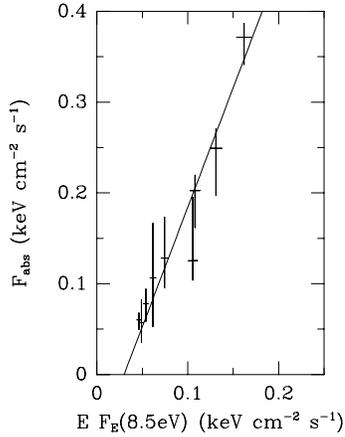}
\end{center}
\caption[]{Comparison of the transmission model prediction with the
$IUE/EXOSAT$ data. The crosses give the flux $F_{abs}$ absorbed by the
cold matter obtained from $EXOSAT$ data, and
$EF_E$(8.5eV) from $IUE$. The solid line gives $EF_E$(8.5eV) as a function
of $F_{abs}$ from the transmission model.} 
\end{figure}

\section{The variability models}

The UV/X-ray correlation can be explained by two specific models with 
reprocessing of X-rays by cold matter (Zdziarski \& Magdziarz 1996). 
The first one is based on reemision 
of the X-ray flux absorbed by Thomson-thin clouds in the line of sight. 
The clouds are dense enough for almost complete termalization. 
The model predicts no Compton reflection which is consistent with the 
broad-band spectra. The second model assumes reprocessing of X-rays and
$\gamma$-rays by a cold, optically-thick accretion disk with dissipative 
patchy corona. The absorbed radiation is reemitted locally in UV as a 
blackbody. The homogeneous corona model is ruled out because the 
hardness of the X-ray spectrum implies that the plasma is photon starved. 
Our second model predicts Compton reflection 
marginally allowed by the observations. Both models satisfy the energy 
balance and provide good fits to the X/$\gamma$-rays and UV data.

The transmission model (Fig.~2) predicts: $F_{UV}=fF_{abs}+F_{0}$ 
where $F_{UV}$ is the integrated UV flux, $F_{0}$ = 0.05 keV cm$^{-2}$ 
s$^{-1}$ is a residual UV flux, and the factor $f$ = 0.60 takes into account
incomplete covering of the X-$\gamma$ source by the clouds as well as 
an efficiency of the absorbed flux conversion into the blackbody continuum. 
We determined the temperature of the absorber as $kT \simeq$ 3 eV 
from the average observed spectral index in the UV between 8.5 and 7.2 eV
($\alpha$ = -0.15; Perola \& Piro 1994). The typical column density of 
a cloud in the partial covering absorber is $N_H \simeq 10^{23}$ cm$^{-2}$ 
with the typical covering factor of $\simeq 0.5$. The size of a single  
cloud is $r_c \simless 10^{7}$. 
The average size of the entire absorber is $\sim 10^{14}$ cm, which satisfy
the limit $r \simless 10^{15}$ cm from the relative UV/X-ray time delay
(Warwick et al.\ 1995). The parameters of the absorber are similar to  
those studied by Ferland \& Rees (1988) and Guilbert \& Rees (1988). 

In the reflection model we assume that all dissipation takes place in the 
corona (e.g. Svensson \& Zdziarski 1994). We integrate 
the blackbody emission over the disk surface with the standard disk 
dissipation rate (Shakura \& Sunyaev 1973). This relates the observed 8.5 eV 
flux to the total UV flux as a function of $r_{S}\mu ^{1/2}$, where $r_S$ is 
the Schwarzschild radius, and $\mu$ is a cosine of the disk inclination angle.
The model predicts $F_{UV}=2(1-A)\mu RF_{X\gamma}$, where A is an albedo 
for the total X-$\gamma$ flux, $F_{X\gamma}$, and $R$ is a ratio of the 
corona emission intercepted by the disk to the luminosity emitted outward. 
Compton reflection for $R <$ 0.5 is consistent with the broad-band spectra. 
We find that all $0 < r_S < 2.4\times 10^{12}$ cm (or equivalently 
$\mu R \simgreat$ 0.15) fit the UV spectral indices (Fig.~3).  
The value of $r_S$ = 1.3 $\times$ $10^{12}$ cm (for $\mu R =$ 0.21) provides 
the best fit to UV fluxes.  

\begin{figure}
\begin{center}
\leavevmode
\epsfysize=7cm\epsfbox[38 30 440 512]{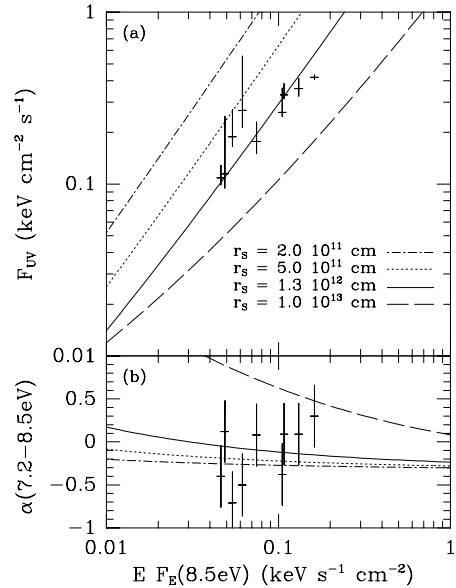}
\end{center}
\caption[]{Comparison of the reflection model predictions with the data.
$(a)$ Crosses give the total UV flux $F_{UV}$ from reemission of the
absorbed X-$\gamma$ flux as obtained extrapolating the $EXOSAT$ X-ray
power laws to $\gamma$-rays assuming thermal Comptonization at $kT$ = 60 keV,
$R$ = 0.5, and $\mu$=0.42 (from {\it HST} observations, Evans et al.\ 1993).
The curves relate the total $F_{UV}$ to $EF_E$(8.5eV), as predicted by the
disk spectrum for various Schwarzchild radii $r_S$. $(b)$ Crosses give the
UV spectral indices (Ulrich et al.\ 1991). Curves give the indices predicted
by the disk-corona model.}
\end{figure}

\end{document}